\documentclass[12pt]{iopart}
\usepackage{latexsym}
\newcommand{\From}{From }       
\newcommand{\bnabla}{\mbox{\boldmath $\nabla$}}

\newcommand{\E}{{\mathcal E}}
\renewcommand{\Re}{\mbox{Re }}
\renewcommand{\Im}{\mbox{Im }}
\newtheorem{definition}{Definition}
\newtheorem{theorem}{Theorem}

\begin{document}

\title{Equatorial symmetry/antisymmetry of stationary axisymmetric
    electrovac spacetimes}
\author{Frederick J.\ Ernst\footnote[1]{e-mail: scientia@localnet.com},
    Vladimir S.\ Manko\footnote[2]{e-mail: vsmanko@fis.cinvestav.mx}
    and Eduardo Ruiz\footnote[3]{e-mail: eruiz@usal.es}}
\address{$\dag$ FJE Enterprises, 511 CR 59, Potsdam, NY 13676 \\
         $\ddag$ Depto.\ de F\'{\i}sica, Centro de Investigaci\'{o}n
     y de Estudios Avanzados del I.P.N., \\
     \hspace{9em} A.P.\ 14-740, 07000 M\'{e}xico, D.F., Mexico \\
     $\S$ Depto.\ de F\'{\i}sica Fundamental, Universidad de Salamanca,
     37008 Salamanca, Spain}

\begin{abstract}
Two theorems are proved concerning how stationary axisymmetric
electrovac spacetimes that are equatorially symmetric or equatorially
antisymmetric can be characterized correctly in terms of the Ernst
potentials $\E$ and $\Phi$ or in terms of axis-data.
\end{abstract}

\section{Introduction}

One of the most significant advances in the construction of physically
interesting equatorially symmetric vacuum spacetimes took place in 1995,
when Kordas \cite{K} and Meinel and Neugebauer \cite{M-N} independently
proved a theorem that allowed one to identify such spacetimes from the
Ernst potentials on the positive $z$-axis, the so-called {\em axis-data},
\textbf{before} the complete solution was constructed.  That this result
was not extended to electrovac spacetimes for over a decade is perhaps a
measure of the difficulty of doing so.

Recently, however, Pach\'{o}n and Sanabria-G\'{o}mez \cite{P-SG} articulated
but did not prove a conjecture concerning the electrovac extension.  One
of the authors of the present paper was suspicious about the validity of
the conjecture as stated, especially because it seemed not to take account
of the fact that the spacetime geometry is not altered when the
electromagnetic Ernst potential $\Phi$ is subjected to a duality rotation.
Finally, we realized that their conjecture is false if one does not replace
the $\pm$ of those authors by our more general coefficient $e^{2i\delta}$.
With this alteration, we were able to develop what we believe is a genuine
proof of the modified Pach\'{o}n--Sanabria-G\'{o}mez conjecture, even
including an extension to the equatorially antisymmetric case.

\begin{definition}
The stationary axisymmetric spacetime with line element
\begin{equation}
ds^{2} = f(z,\rho)^{-1}\left[e^{2\gamma(z,\rho)}(dz^{2}+d\rho^{2})
+\rho^{2}d\varphi^{2}\right]-f(z,\rho)[dt-\omega(z,\rho)d\varphi]^{2}
\label{ds2}
\end{equation}
will be said to be ``equatorially symmetric'' when there exists a choice
of the origin such that under the transformation
\begin{equation}
z\rightarrow z'=-z, \quad \rho \rightarrow \rho'=\rho,
\label{transf}
\end{equation}
the metrical fields $f$, $\omega$ and $\gamma$ are all even functions
of $z$, and will be said to be ``equatorially antisymmetric" when there
exists a choice of the origin and a constant $\omega_{0}$ such that
under the transformation (\ref{transf}), the metrical fields $f$ and
$\gamma$ are even functions of $z$, while $\omega-\omega_{0}$ is an
odd function of $z$.
\end{definition}

\section{Statement and proof of fundamental theorem}

\begin{theorem}
A stationary axisymmetric electrovac solution of the Ernst equations
\begin{eqnarray}
(\Re\E + |\Phi|^{2}) \nabla^{2}\E & = &
(\bnabla\E + 2 \Phi^{*}\bnabla\Phi) \cdot \bnabla\E,
\label{elec1} \\
(\Re\E + |\Phi|^{2}) \nabla^{2}\Phi & = &
(\bnabla\E + 2 \Phi^{*}\bnabla\Phi) \cdot \bnabla\Phi,
\label{elec2}
\end{eqnarray}
is equatorially symmetric if and only if
\begin{equation}
\E(-z,\rho) = \E(z,\rho)^{*},\quad\Phi(-z,\rho) =
e^{2i\delta}\Phi(z,\rho)^{*},
\label{eqsym}
\end{equation}
where $\delta$ is a real constant, and is equatorially antisymmetric if
and only if
\begin{equation}
\E(-z,\rho) = \E(z,\rho),\quad \Phi(-z,\rho) = \pm\Phi(z,\rho).
\label{eqasym}
\end{equation}
\end{theorem}
In the Ernst equations the symbol $\nabla^{2}$ denotes the Laplacian
and $\bnabla A\cdot\bnabla B$ denotes a scalar product of gradients in a
three-dimensional axisymmetric flat space.

\textbf{Proof:}
Recall that the metrical fields $f(z,\rho)$, $\omega(z,\rho)$ and
$\gamma(z,\rho)$ are obtained from the Ernst potentials $\E(z,\rho)$
and $\Phi(z,\rho)$ using the equations
\begin{equation}
f = \Re{\E}+|\Phi|^{2}, \quad \chi = \Im{\E},
\label{fchi}
\end{equation}
\begin{equation}
\omega_{,z} = -\rho f^{-2}[\chi_{,\rho}+2\Im(\Phi^{*}\Phi_{,\rho})], \;
\omega_{,\rho} = \rho f^{-2}[\chi_{,z}+2\Im(\Phi^{*}\Phi_{,z})],
\label{omega}
\end{equation}
and
\begin{eqnarray}
\gamma_{,z} & = & \frac{1}{4}\rho f^{-2}[
(\E_{,\rho}+2\Phi^{*}\Phi_{,\rho})(\E_{,z}^{*}+2\Phi\Phi_{,z}^{*})
\nonumber \\ & & \mbox{ }
+ (\E_{,z}+2\Phi^{*}\Phi_{,z})
(\E_{,\rho}^{*}+2\Phi\Phi_{,\rho}^{*})]
\label{gamz} \\ & & \mbox{ }
- \rho f^{-1}(\Phi_{,\rho}\Phi_{,z}^{*}+\Phi_{,z}\Phi_{,\rho}^{*}),
\nonumber \\
\gamma_{,\rho} & = & \frac{1}{4}\rho f^{-2}[
(\E_{,\rho}+2\Phi^{*}\Phi_{,\rho}) (\E_{,\rho}^{*}+2\Phi\Phi_{,\rho}^{*})
\nonumber \\ & & \mbox{ }
\mbox{ } - (\E_{,z}+2\Phi^{*}\Phi_{,z}) (\E_{,z}^{*}+2\Phi\Phi_{,z}^{*})]
\label{gamrho} \\ & &
- \rho f^{-1}(\Phi_{,\rho}\Phi_{,\rho}^{*}-\Phi_{,z}\Phi_{,z}^{*}).
\nonumber
\end{eqnarray}
The integrability of (\ref{omega}) and of (\ref{gamz}) and
(\ref{gamrho}) is guaranteed by the Ernst equations (\ref{elec1}) and
(\ref{elec2}).  When we suppress the arguments of functions such as
$\omega_{,z}$ and $\omega_{,\rho}$, it is to be understood that we mean
those functions, the values of which are values of the derivatives of the
function $\omega$ with respect to its first argument $z$ and its second
argument $\rho$, respectively.

\From (\ref{eqsym}), one can use (\ref{fchi}) to show that
$f(-z,\rho)=f(z,\rho)$ and $\chi(-z,\rho)=-\chi(z,\rho)$, and
(by choosing the integration constant appropriately) one can use
(\ref{omega}) to show that $\omega(-z,\rho)=\omega(z,\rho)$.  Finally,
(by choosing the integration constant appropriately) one can use
(\ref{gamz}) and (\ref{gamrho}) to show that
$\gamma(-z,\rho)=\gamma(z,\rho)$, thus establishing equatorial symmetry.
In a similar manner, starting with (\ref{eqasym}), one can establish
equatorial antisymmetry.

In order to prove, conversely, that equatorial symmetry implies
(\ref{eqsym}) and equatorial antisymmetry implies (\ref{eqasym}), we
note that all the diagonal components of the Ricci tensor $R_{ij}$
with respect to the orthonormal tetrad
\begin{eqnarray}
e^{1} & = & f^{-1/2}e^{\gamma}d\rho, \\
e^{2} & = & f^{-1/2}e^{\gamma}dz, \\
e^{3} & = & f^{-1/2}\rho d\phi, \\
e^{4} & = & f^{1/2}(dt-\omega d\phi),
\end{eqnarray}
are even functions of $z$ both in the equatorially symmetric case and
the equatorially antisymmetric case, while $R_{12}$ is an odd function
of $z$ in both cases.  On the other hand, $R_{34}$ is even in the
equatorially symmetric case and odd in the equatorially antisymmetric
case.  All other components of the Ricci tensor vanish identically
This implies similar behaviour of the orthonormal components of the
stress-energy tensor $T_{ij}$, from which the following {\em quadratic
constraints} can be inferred:
\begin{enumerate}
\item
$|\Phi_{,z}|^{2}$ is an even function of $z$,
\item
$|\Phi_{,\rho}|^{2}$ is an even function of $z$,
\item
$\Im(\Phi_{,z}\Phi_{,\rho}^{*})$ is an even function of $z$
in the symmetric case, and an odd function of $z$ in the
antisymmetric case,
\item
$\Re(\Phi_{,z}\Phi_{,\rho}^{*})$ is an odd function of $z$.
\end{enumerate}
These results follow directly from a paper of Ernst \cite{E} that
was published in 1974. There a complex null tetrad that is simply
related to the above orthonormal tetrad was employed.\footnote{See
the website http://members.localnet.com/$\sim$atheneum/gs/gs.html
for the source code and output of one symbolic manipulation
program in which the orthonormal tetrad components of the Ricci
tensor are evaluated in terms of the metric fields $f$, $\omega$
and $\gamma$ and their derivatives, and another program in which
the orthonormal tetrad components of the stress-energy tensor are
evaluated in terms of the complex potential $\Phi$.}

The first two quadratic constraints can be expressed as
\begin{eqnarray}
|\Phi_{,z}(-z,\rho)| & = & |\Phi_{,z}(z,\rho)|, \label{R1}\\
|\Phi_{,\rho}(-z,\rho)| & = & |\Phi_{,\rho}(z,\rho)|, \label{R2}
\end{eqnarray}
while the third and fourth constraints can be expressed as
\begin{equation}
\Phi_{,z}(-z,\rho)\Phi^{*}_{,\rho}(-z,\rho) =
-[\Phi_{,z}(z,\rho)\Phi^{*}_{,\rho}(z,\rho)]^{*} \label{Rsym}
\end{equation}
in the symmetric case and as
\begin{equation}
\Phi_{,z}(-z,\rho)\Phi^{*}_{,\rho}(-z,\rho) =
-\Phi_{,z}(z,\rho)\Phi^{*}_{,\rho}(z,\rho) \label{Rasym}
\end{equation}
in the antisymmetric case.

We note in passing that, using the chain rule, all equations can be
rewritten in terms of
\begin{eqnarray}
[\Phi(-z,\rho)]_{,z} & := &
\frac{\partial}{\partial z}\left[\Phi(-z,\rho)\right]
= -\Phi_{,z}(-z,\rho), \\ \mbox{ }
[\Phi(-z,\rho)]_{,\rho} & := &
\frac{\partial}{\partial \rho}\left[\Phi(-z,\rho)\right]
= \Phi_{,\rho}(-z,\rho),
\end{eqnarray}
if one wishes to extend the use of comma notation to partial derivatives in
general.  In this case, to avoid ambiguity, one is obliged never to suppress
the arguments of the function to which partial differentiation is applied.

Without loss of generality we may write
\begin{equation}
\Phi_{,z}(z,\rho) = e^{i\alpha(z,\rho)}h_{1}(z,\rho), \;
\Phi_{,\rho}(z,\rho) = e^{i\beta(z,\rho)}h_{2}(z,\rho),
\label{alphabeta}
\end{equation}
and
\begin{equation}
\Phi^{*}_{,z}(z,\rho) = e^{-i\alpha(z,\rho)}h_{1}(z,\rho), \;
\Phi^{*}_{,\rho}(z,\rho) = e^{-i\beta(z,\rho)}h_{2}(z,\rho),
\end{equation}
where $\alpha$, $\beta$, $h_{1}$ and $h_{2}$ are all real functions
of $(z,\rho)$.

By (\ref{R1}) and (\ref{R2}), $h_{1}$ and $h_{2}$ have definite
parity, either even or odd.  In the symmetric case, (\ref{Rsym})
tells us that
\begin{equation}
e^{i[\alpha(-z,\rho)-\beta(-z,\rho)]}h_{1}(-z,\rho)h_{2}(-z,\rho) =
- e^{-i[\alpha(z,\rho)-\beta(z,\rho)]}h_{1}(z,\rho)h_{2}(z,\rho),
\end{equation}
and in the antisymmetric case, (\ref{Rasym}) tells us that
\begin{equation}
e^{i[\alpha(-z,\rho)-\beta(-z,\rho)]}h_{1}(-z,\rho)h_{2}(-z,\rho) =
- e^{i[\alpha(z,\rho)-\beta(z,\rho)]}h_{1}(z,\rho)h_{2}(z,\rho).
\end{equation}
These two equations can also be expressed in the forms
\begin{equation}
h_{1}(-z,\rho)h_{2}(-z,\rho) = -
e^{-2i[\alpha_{+}(z,\rho)-\beta_{+}(z,\rho)]}h_{1}(z,\rho)h_{2}(z,\rho)
\end{equation}
and
\begin{equation}
h_{1}(-z,\rho)h_{2}(-z,\rho) = -
e^{2i[\alpha_{-}(z,\rho)-\beta_{-}(z,\rho)]}h_{1}(z,\rho)h_{2}(z,\rho),
\end{equation}
respectively, where
\begin{equation}
\alpha_{\pm}(z,\rho) := \frac{1}{2}[\alpha(z,\rho) \pm \alpha(-z,\rho)], \;
\beta_{\pm}(z,\rho) := \frac{1}{2}[\beta(z,\rho) \pm \beta(-z,\rho)],
\end{equation}
are the even and odd parts of $\alpha(z,\rho)$ and $\alpha(-z,\rho)$
and of $\beta(z,\rho)$ and $\beta(-z,\rho)$, respectively.

Whatever may be the parities of $h_{1}(z,\rho)$ and $h_{2}(z,\rho)$,
one can show that equatorial symmetry implies
\begin{equation}
\Phi_{,z}(-z,\rho) = \mp e^{2i\beta_{+}(z,\rho)}\Phi_{,z}(z,\rho)^{*},
\;
\Phi_{,\rho}(-z,\rho) = \pm e^{2i\beta_{+}(z,\rho)}\Phi_{,\rho}(z,\rho)^{*},
\label{zrhosym}
\end{equation}
while equatorial antisymmetry implies
\begin{equation}
\Phi_{,z}(-z,\rho) = \mp e^{-2i\beta_{-}(z,\rho)}\Phi_{,z}(z,\rho),
\;
\Phi_{,\rho}(-z,\rho) = \pm e^{-2i\beta_{-}(z,\rho)}\Phi_{,\rho}(z,\rho).
\label{zrhoasym}
\end{equation}
If we can establish that $\beta_{+}(z,\rho)$ and $\beta_{-}(z,\rho)$
are constants, then we shall be able to conclude that
\begin{eqnarray}
\Phi(-z,\rho) & = & \pm e^{-2i\beta_{+}}\Phi(z,\rho)^{*}
\mbox{ in the symmetric case, and }
\label{symcase} \\
\Phi(-z,\rho) & = & \pm \Phi(z,\rho)
\mbox{ in the antisymmetric case.}
\label{asymcase}
\end{eqnarray}

Noting that both $\beta_{+}$ and $\beta_{-}$ are real functions of $(z,\rho)$,
we may conclude from (\ref{zrhosym}) and (\ref{zrhoasym}) that
\begin{equation}
\Phi_{,\rho} \beta_{\pm,z} - \Phi_{,z} \beta_{\pm,\rho} = 0.
\label{one}
\end{equation}

Writing the Ernst equation for $\Phi$ and the complex conjugate of that
equation in the respective forms
\begin{eqnarray}
f(z,\rho)[\Phi_{,z,z}(z,\rho)+\Phi_{,\rho,\rho}(z,\rho) +
\rho^{-1}\Phi_{,\rho}(z,\rho)] & & \nonumber \\ \mbox{ }
- [f_{,z}(z,\rho)-i\rho^{-1}f(z,\rho)^{2}\omega_{,\rho}(z,\rho)]
\Phi_{,z}(z,\rho) \nonumber \\ \mbox{ }
+ [f_{,\rho}(z,\rho)+i\rho^{-1}f(z,\rho)^{2}\omega_{,z}(z,\rho)]
\Phi_{,\rho}(z,\rho)
& = & 0
\label{plus2}
\end{eqnarray}
and
\begin{eqnarray}
f(z,\rho)[\Phi^{*}_{,z,z}(z,\rho)+\Phi^{*}_{,\rho,\rho}(z,\rho) +
\rho^{-1}\Phi^{*}_{,\rho}(z,\rho)] & & \nonumber \\ \mbox{ }
- [f_{,z}(z,\rho)+i\rho^{-1}f(z,\rho)^{2}\omega_{,\rho}(z,\rho)]
\Phi^{*}_{,z}(z,\rho) \nonumber \\ \mbox{ }
+ [f_{,\rho}(z,\rho)-i\rho^{-1}f(z,\rho)^{2}\omega_{,z}(z,\rho)]
\Phi^{*}_{,\rho}(z,\rho)
& = & 0,
\label{plus}
\end{eqnarray}
we can infer that, in the equatorially symmetric case,
\begin{eqnarray}
f(z,\rho)[\Phi_{,z,z}(-z,\rho)+\Phi_{,\rho,\rho}(-z,\rho) +
\rho^{-1}\Phi_{,\rho}(-z,\rho)] & & \nonumber \\ \mbox{ }
+ [f_{,z}(z,\rho)+i\rho^{-1}f(z,\rho)^{2}\omega_{,\rho}(z,\rho)]
\Phi_{,z}(-z,\rho) \nonumber \\ \mbox{ }
+ [f_{,\rho}(z,\rho)-i\rho^{-1}f(z,\rho)^{2}\omega_{,z}(z,\rho)]
\Phi_{,\rho}(-z,\rho)
& = & 0,
\label{minus}
\end{eqnarray}
while, in the equatorially antisymmetric case,
\begin{eqnarray}
f(z,\rho)[\Phi_{,z,z}(-z,\rho)+\Phi_{,\rho,\rho}(-z,\rho) +
\rho^{-1}\Phi_{,\rho}(-z,\rho)] & & \nonumber \\ \mbox{ }
+ [f_{,z}(z,\rho)-i\rho^{-1}f(z,\rho)^{2}\omega_{,\rho}(z,\rho)]
\Phi_{,z}(-z,\rho) \nonumber \\ \mbox{ }
+ [f_{,\rho}(z,\rho)+i\rho^{-1}f(z,\rho)^{2}\omega_{,z}(z,\rho)]
\Phi_{,\rho}(-z,\rho)
& = & 0.
\label{minus2}
\end{eqnarray}

Into (\ref{minus}) we can substitute (\ref{zrhosym}) and
\begin{eqnarray}
\Phi_{,z,z}(-z,\rho) & = & \pm e^{2i\beta_{+}(z,\rho)}
[\Phi^{*}_{,z,z}(z,\rho)+2i\beta_{+,z}(z,\rho)\Phi^{*}_{,z}(z,\rho)], \\
\Phi_{,\rho,\rho}(-z,\rho) & = & \pm e^{2i\beta_{+}(z,\rho)}
[\Phi^{*}_{,\rho,\rho}(z,\rho)
+2i\beta_{+,\rho}(z,\rho)\Phi^{*}_{,\rho}(z,\rho)],
\end{eqnarray}
and then use the resulting equation together with (\ref{plus})
to show that
\begin{equation}
\Phi^{*}_{,z}\beta_{+,z} + \Phi^{*}_{,\rho}\beta_{+,\rho} = 0.
\end{equation}
Similarly, into (\ref{minus2}) we can substitute
(\ref{zrhoasym}) and
\begin{eqnarray}
\Phi_{,z,z}(-z,\rho) & = & \pm e^{-2i\beta_{-}(z,\rho)}
[\Phi^{*}_{,z,z}(z,\rho)-2i\beta_{-,z}(z,\rho)\Phi^{*}_{,z}(z,\rho)], \\
\Phi_{,\rho,\rho}(-z,\rho) & = & \pm e^{-2i\beta_{-}(z,\rho)}
[\Phi^{*}_{,\rho,\rho}(z,\rho)
-2i\beta_{-,\rho}(z,\rho)\Phi^{*}_{,\rho}(z,\rho)],
\end{eqnarray}
and then use the resulting equation together with (\ref{plus2})
to show that
\begin{equation}
\Phi_{,z}\beta_{-,z} + \Phi_{,\rho}\beta_{-,\rho} = 0.
\end{equation}
In either case, because of the reality of $\beta_{\pm}$, we have
\begin{equation}
\Phi^{*}_{,z}\beta_{\pm,z} + \Phi^{*}_{,\rho}\beta_{\pm,\rho} = 0.
\label{two}
\end{equation}

\From (\ref{one}) and (\ref{two}) we infer that
\begin{equation}
\left(\begin{array}{cc}
\Phi_{,\rho} & -\Phi_{,z} \\
\Phi^{*}_{,z} & \Phi^{*}_{,\rho}
\end{array} \right) \left( \begin{array}{c}
\beta_{\pm,z} \\ \beta_{\pm,\rho}
\end{array} \right) = \left( \begin{array}{c}
0 \\ 0
\end{array} \right).
\end{equation}
In the case of a vacuum spacetime, where $\Phi=0$, (\ref{eqsym})
and (\ref{eqasym}) can be established easily.  If we are not dealing
with a vacuum, then $|\Phi_{,\rho}|^{2}+|\Phi_{,z}|^{2} > 0$, so the
square matrix is invertible, and we conclude that
$\beta_{\pm,z} = \beta_{\pm,\rho} = 0$ and the $\beta_{\pm}$ are
constants.  In particular, $\beta_{-}=0$.

At this point we turn our attention to (\ref{fchi}) and (\ref{omega}),
which can be shown to imply (\ref{eqsym}) in the equatorially
symmetric case and (\ref{eqasym}) in the equatorially antisymmetric
case.
\textbf{QED}

\section{A theorem concerning axis data}

Using Theorem 1 we can now establish the following extension of the 1995
result of Kordas, Meinel and Neugebauer:
\begin{theorem}
An asymptotically flat electrovac solution constructed from data
$e_{+}(z) = \E(z,0)$ and $f_{+}(z) = \Phi(z,0)$ specified on a connected
open interval extending to $z = +\infty$ on the positive $z$-axis will
be equatorially symmetric if and only if
\begin{equation}
e_{+}(z)[e_{+}(-z)]^{*}=1 \mbox{ and }
f_{+}(z) = - e^{2i\delta}[f_{+}(-z)]^{*}e_{+}(z),
\label{sym}
\end{equation}
and will be equatorially antisymmetric if and only if
\begin{equation}
e_{+}(z)e_{+}(-z)=1 \mbox{ and }
f_{+}(z) = \mp f_{+}(-z)e_{+}(z),
\label{asym}
\end{equation}
where $e_{+}(z)$ and $f_{+}(z)$ have both been analytically
extended from the positive $z$-axis to the negative $z$-axis, and
$\delta$ is a real constant.
\end{theorem}
Please note that the analytically extended axis-data $e_{+}(z)$ and
$f_{+}(z)$ are not in general, for negative values of $z$, equal to
the axis values $\E(z,0)$ and $\Phi(z,0)$ of the Ernst potentials of
the solution constructed from that axis data!

\textbf{Proof:}
It is useful to reexpress (\ref{eqsym}) and (\ref{eqasym}) of
Theorem 1 in terms of the auxiliary complex potentials \cite{S-A}
\begin{equation}
\xi = \frac{1-\E}{1+\E}, \quad \eta = \frac{2\Phi}{1+\E}.
\label{xieta}
\end{equation}
In the equatorially symmetric case
\begin{equation}
\xi(-z,\rho) = \xi(z,\rho)^{*}, \quad
\eta(-z,\rho) = e^{2i\delta}\eta(z,\rho)^{*},
\end{equation}
while in the equatorially antisymmetric case
\begin{equation}
\xi(-z,\rho) = \xi(z,\rho), \quad
\eta(-z,\rho) = \pm\eta(z,\rho).
\end{equation}
We also introduce the auxiliary potentials $\tilde{\xi}(\bar{z},\bar{\rho})$
and $\tilde{\eta}(\bar{z},\bar{\rho})$, where
\begin{equation}
\tilde{\xi}(\bar{z},\bar{\rho}) := \sqrt{z^{2}+\rho^{2}}\;\xi(z,\rho), \quad
\tilde{\eta}(\bar{z},\bar{\rho}) := \sqrt{z^{2}+\rho^{2}}\;\eta(z,\rho),
\end{equation}
and
\begin{equation}
\bar{z} := \frac{z}{z^{2}+\rho^{2}}, \quad
\bar{\rho} := \frac{\rho}{z^{2}+\rho^{2}}.
\end{equation}
In terms of these potentials, (\ref{eqsym}) assume the form
\begin{equation}
\tilde{\xi}(-\bar{z},\bar{\rho}) = \tilde{\xi}(\bar{z},\bar{\rho})^{*},
\quad
\tilde{\eta}(-\bar{z},\bar{\rho}) = \epsilon e^{2i\delta}
\tilde{\eta}(\bar{z},\bar{\rho})^{*},
\label{tilxieta}
\end{equation}
and (\ref{eqasym}) assume the form
\begin{equation}
\tilde{\xi}(-\bar{z},\bar{\rho}) = \tilde{\xi}(\bar{z},\bar{\rho}),
\quad
\tilde{\eta}(-\bar{z},\bar{\rho}) = \epsilon\tilde{\eta}(\bar{z},\bar{\rho}),
\label{atilxieta}
\end{equation}
where $\epsilon = \pm 1$ and the complex coefficients $m_{n}$ and $q_{n}$
in the power series expansions
\begin{equation}
\tilde{\xi}(\bar{z},0) = \sum_{n=0}^{\infty}m_{n}\bar{z}^{n}, \quad
\tilde{\eta}(\bar{z},0) = \sum_{n=0}^{\infty}q_{n}\bar{z}^{n}
\label{moments}
\end{equation}
are directly related to the various gravitational and electromagnetic
multipole moments, respectively.

\From (\ref{tilxieta}) and (\ref{moments}) it can easily be
seen that, in the equatorially symmetric case, the coefficients $m_{n}$
must be real for $n$ even and imaginary for $n$ odd, while from
(\ref{atilxieta}) and (\ref{moments}) it follows that, in the equatorially
antisymmetric case, the coefficients $m_{n}$ must vanish for $n$ odd.

On the other hand, the nature of the coefficients $q_{n}$ depends upon
the sign of $\epsilon$.  When $\epsilon = 1$, the nature of the $q_{n}$
is identical to that prescribed in the last paragraph for the $m_{n}$.
When $\epsilon = -1$, the coefficients $q_{n}$ must be imaginary for
$n$ even and real for $n$ odd in the equatorially symmetric case, and
the coefficients $q_{n}$ must vanish for $n$ even in the equatorially
antisymmetric case.  Of course, without altering the spacetime geometry
one can subject $\Phi$ to a duality rotation
\begin{equation}
\Phi \rightarrow e^{i\delta}\Phi,
\end{equation}
as a result of which the $q_{n}$ acquire a common constant complex phase
$e^{i\delta}$.

We shall apply the method of reasoning developed by Kordas, using the
analytic character of $\tilde{\xi}(\bar{z},0)$ and $\tilde{\eta}(\bar{z},0)$
in a sufficiently small open neighborhood of $\bar{z}=0$.  In the
equatorially symmetric case, for $z>0$ we have
\begin{eqnarray}
e_{+}(z) & = & \frac{1-\xi(z,0)}{1+\xi(z,0)}
= \frac{1-\bar{z}\tilde{\xi}(\bar{z},0)}{1+\bar{z}\tilde{\xi}(\bar{z},0)}, \\
f_{+}(z) & = & \frac{\eta(z,0)}{1+\xi(z,0)}
= \frac{\bar{z}\tilde{\eta}(\bar{z},0)}{1+\bar{z}\tilde{\xi}(\bar{z},0)}.
\end{eqnarray}
Analytically extending this to negative values of $z$, we obtain
\begin{eqnarray}
e_{+}(-z) & = &
\frac{1+\bar{z}\tilde{\xi}(-\bar{z},0)}{1-\bar{z}\tilde{\xi}(-\bar{z},0)} =
\frac{1+\bar{z}\tilde{\xi}(\bar{z},0)^{*}}
{1-\bar{z}\tilde{\xi}(\bar{z},0)^{*}}, \\
f_{+}(-z) & = &
-\frac{\bar{z}\tilde{\eta}(-\bar{z},0)}{1-\bar{z}\tilde{\xi}(-\bar{z},0)}
= - e^{2i\delta}\frac{\bar{z}\tilde{\eta}(\bar{z},0)^{*}}
{1-\bar{z}\tilde{\xi}(\bar{z},0)^{*}},
\end{eqnarray}
where we have used (\ref{tilxieta}).  Therefore, the solution will
be equatorially symmetric if and only if (\ref{sym}) hold.  The
proof that the solution will be equatorially antisymmetric if and only
if (\ref{asym}) hold proceeds in the same way.
\textbf{QED}

\section{Concluding remarks}

One should note that (\ref{sym}) and (\ref{asym}) of Theorem 2
apply only to asymptotically flat spacetimes (up to the NUT parameter),
whereas (\ref{eqsym}) and (\ref{eqasym}) of Theorem 1 apply more
generally.

An example of an equatorially symmetric electrovac solution is the
electrically and magnetically charged Kerr \cite{K-N} solution,
while an equatorially symmetric static electrovac solution is the
magnetic dipole solution of Gutsunaev and Manko \cite{G-M}.  Finally,
an example of an equatorially antisymmetric electrovac solution is the
binary system of `antisymmetric' Kerr--Newman masses of Bret\'{o}n
and Manko \cite{B-M}.  We hope that the theorems proved in this paper
will aid in the construction of many other solutions of physical interest.

\ack
We would like to thank Leonardo Pach\'on and Jos\'e Sanabria--G\'omez
for sending us a preprint of their paper, and to thank anonymous
referees for suggesting useful amendments to the manuscript of the
present paper.  This work was partially supported by Project 45946--F
from CONACyT of Mexico and by Project BFM2003--02121 from MCyT of Spain.


\section*{References}
\begin{thebibliography}{10}
\bibitem{K}
Kordas~P 1995 Reflection-symmetric, asymptotically flat solutions
of the vacuum axistationary Einstein equations,  {\it Class.\
Quantum Grav.}\ \textbf{12} 2037
\bibitem{M-N}
Meinel~R and Neugebauer~G 1995 Asymptotically flat solutions to
the Ernst equation with reflection symmetry, {\it Class.\ Quantum
Grav.}\ \textbf{12} 2045
\bibitem{P-SG}
Pach\'{o}n~L~A and Sanabria-G\'{o}mez~J~D 2006 Note on reflection
symmetry in stationary axisymmetric electrovacuum space-times,
{\it Class.\ Quantum Grav.}\ \textbf{23} 3251
\bibitem{E}
Ernst~F~J 1974 Complex potential formulation of the axially
symmetric gravitational field problem, {\it J.\ Math.\ Phys.}\
\textbf{15} 1409
\bibitem{S-A}
Sotiriou~T~P and Apostolatos~T~A 2004 Corrections and comments on
the multipole moments of axisymmetric electrovacuum spacetimes,
{\it Class.\ Quantum Grav.}\ \textbf{21} 5727
\bibitem{K-N}
Newman~E~T, Couch~E, Chinnapared~K, Exton~A, Prakash~A and
Torrence~R 1965 Metric of a rotating, charged mass, {\it J.\
Math.\ Phys.}\ \textbf{6} 918
\bibitem{G-M}
Gutsunaev~Ts~I and Manko~V~S 1988 On a family of solutions of the
Einstein--Maxwell equations, {\it Gen.\ Relat.\ Grav.}\
\textbf{20} 4. Multiply (47) by $i$ to get our $\Phi$.
Equation (48) gives $f$.
\bibitem{B-M}
Bret\'{o}n~N and Manko~V~S 1995 A binary system of `antisymmetric'
Kerr--Newman masses, {\it Class.\ Quantum Grav.}\ \textbf{12} 1969
\end{thebibliography}
\end{document}